\documentstyle[twoside,fleqn,espcrc2,epsfig]{article}

\newcommand{\Cf}{$^{252}\mbox{Cf}$}
\newcommand{\CoF}{$^{57}\mbox{Co}$}
\newcommand{\CoS}{$^{60}\mbox{Co}$}
\newcommand{\Pb}{$^{210}\mbox{Pb}$}

\newcommand{\SPEC}{\address{CEA, Centre d'Etudes de Saclay, 
DSM/DRECAM, Service de Physique de l'Etat Condens\'e, \\
F-91191 Gif-sur-Yvette Cedex, France}}
\newcommand{\SPP}{\address{CEA, Centre d'Etudes de Saclay, 
DSM/DAPNIA, Service de Physique des Particules, \\
F-91191 Gif-sur-Yvette Cedex, France}}
\newcommand{\CSNSM}{\address{Centre de Spectrom\'etrie Nucl\'eaire 
et Spectrom\'etrie de Masse, IN2P3-CNRS, Univ. Paris XI, Bat. 108, \\
F-91405 Orsay Cedex, France}}
\newcommand{\IPNL}{\address{Institut de Physique Nucl\'eaire de Lyon 
and Universit\'e Claude Bernard, Lyon I, IN2P3-CNRS, \\
43 Bd. du 11 novembre 1918, F-69622 Villeurbanne Cedex, France}}
\newcommand{\IAP}{\address{Institut d'Astrophysique de Paris, INSU-CNRS, 
98bis Bd. Arago, F-75014 Paris, France}}
\newcommand{\LSM}{\address{Laboratoire Souterrain de Modane, CEA-CNRS, 
90 Rue Polset, F-73500 Modane, France}}

\newcommand{\CSNSMbis}{\hbox{$^{\rm a}$}} 
\newcommand{\IPNLbis}{\hbox{$^{\rm b}$}} 
\newcommand{\SPECbis}{\hbox{$^{\rm c}$}} 
\newcommand{\SPPbis}{\hbox{$^{\rm d}$}} 
 
\newcommand{\IAPbis}{\hbox{$^{\rm f}$}} 

\title{Status of the EDELWEISS Experiment}
 
\author{
{\large Presented by Philippe  Di Stefano, CEA/Saclay, DSM/DAPNIA/SPP\\
\vspace{0.3cm}}
L.~Berg\'e\CSNSM, 
I.~Berkes\IPNL, 
B.~Chambon\IPNLbis, 	
M.~Chapellier\SPEC, 
G.~Chardin\SPP, 
P.~Charvin\LSM, 
M.~De~J\'esus\IPNLbis,
P.~Di~Stefano\SPPbis,
D.~Drain\IPNLbis, 
L.~Dumoulin\CSNSMbis, 
C.~Goldbach\IAP, 
A.~Juillard\CSNSMbis, 
D.~L'H\^ote\SPECbis, 
S.~Marnieros\CSNSMbis, 
L.~Miramonti\SPPbis, 
L.~Mosca\SPPbis,	
X.-F.~Navick\SPECbis, 
G.~Nollez\IAPbis, 
P.~Pari\SPECbis, 
C.~Pastor\IPNLbis, 
S.~Pecourt\IPNLbis,
R.~Tourbot\SPECbis, 
D.~Yvon\SPPbis
}

\begin{document}

\begin{abstract}
The	status of the EDELWEISS	experiment (underground dark matter search with 
heat-ionisation bolometers)
is reviewed.	 
Auspicious results achieved with 
a prototype 
70~g Ge	heat-ionisation	detector under a 2 V reverse bias tension 
are discussed. Based on gamma and neutron 
calibrations, a best-case rejection factor, over the 15-45~keV	range, of 99.7\%	for
gammas,	with an	acceptance of 94\% for neutrons, is	presented first.
Some operational results of	physical interest obtained under poor 
low radioactivity conditions follow.  They include a 
raw event rate
of around 30~events/day/kg/keV over the same	energy range,
and, after rejection of part of the background, lead to	a 
conservative
upper limit	on the
signal of approximately	1.6~events/day/kg/keV at a 90\%	confidence
level.	Performance	degrading surface effects of the detector are
speculated upon; and planned upgrades are summarized.
\end{abstract}

\maketitle

\section{DARK MATTER AND DIRECT DETECTION}
Ironically enough, the problem of dark matter has been glaring in physics for several 
years now \cite{proc:Schramm,art:Mosca}, and the case for a 
non-baryonic contribution to it is compelling \cite{proc:Bergstrom}.  Moreover, structure formation 
suggests
 a sizable amount of cold dark matter made up of particles in 
the GeV to TeV range.  
Various supersymmetric
theories generate a plausible 
candidate for WIMPs, the Lightest Supersymmetric Particle, 
which would be stable in the case of R-parity conservation 
\cite{art:Jungman}.  Several searches for these WIMPs are underway, using either 
indirect means (i.e. looking for products of WIMP annihilation) or 
direct ones \cite{proc:Caldwell}.  The latter, first 
suggested by Goodman and Witten \cite{art:Goodman}, seek interactions of 
WIMPs themselves in detectors, which historically have been 
semi-conducting devices, scintillators and, more recently, 
bolometers.  Bolometers are attractive for several reasons : they 
achieve excellent energy resolutions and low thresholds, and can be fashioned from various 
materials. Furthermore, if made of $\mbox{Ge}$ or $\mbox{Si}$, they should be easily cleansed of radioactive impurities, and 
can reject 
 a large fraction
of background noise thanks to the possibility of simultaneous heat and 
ionisation measurement \cite{art:Shutt3}.  The EDELWEISS experiment, 
presented hereafter, has developed and tested such bolometers.

\section{THE EDELWEISS EXPERIMENT AT THE LABORATOIRE SOUTERRAIN DE 
MODANE}
EDELWEISS (Exp\'erience pour DEtecter Les WIMPs En SIte Souterrain) is 
an underground direct WIMP detection experiment.  The underground 
site is the Laboratoire Souterrain de Modane (LSM), located by a highway 
tunnel beneath the Franco-Italian Alps.  The mountain provides protection 
equivalent to 4800~meters of water, which yields a six order of 
magnitude reduction in the cosmic muon flux.  Natural radioactivity of 
the surrounding rock generates $4.10^{-6} \mbox{ /s/cm}^{2}$ worth of fast 
neutrons.  The importance of this value lies in the fact that 
neutrons, which interact \emph{a priori} like WIMPs, are expected to 
make up the ultimate background noise for bolometers.

The initial experimental setup has been described in 
Ref.~\cite{art:deBellefon}.  Since then, several enhancements have been made.  Specifically, 
the dilution refrigerator now reaches temperatures as low 
as 10 mK, 
with a cooling power of 
100~$\mu$W 
at 100~mK; the electronics have 
undergone a straightforward 
upgrade since Ref.~\cite{art:Yvon}; and a new detector has been tested.  For the 
purpose of these tests and data taking, the part of the low radioactivity shielding 
closest to the detector has been omitted, and no radon removal has 
been performed.  Moreover, off-line data analysis is now performed 
using optimal filter techniques in the frequency domain 
\cite{art:Moseley}.

\section{DETECTOR AND CALIBRATION}
\begin{figure}[ht]
	\centering
	\epsfig{file=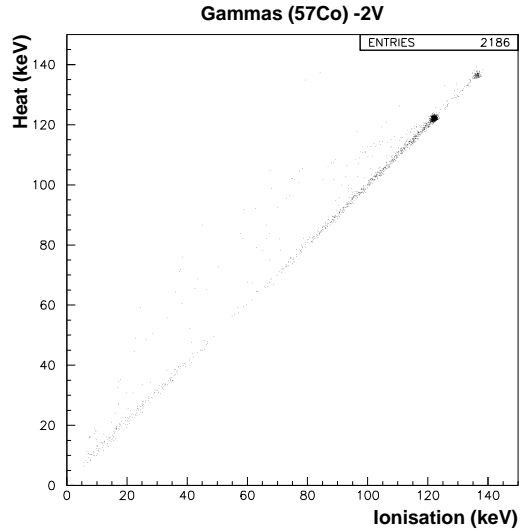,width=\linewidth}
	\caption{Energy calibration of detector using a \CoF\ source.  
	Energies are normalized to equivalent electron recoils.  Some 
	off-axis events, including drooling from the 122 keV peak, are 
	visible. Smaller 136 keV sibling appears also.}
	\label{fig:Calibr}
\end{figure}

Detector design is detailed in Ref.~\cite{art:EdelwDet}. Performances are summarized in 
Table~\ref{tab:DetectorPerformance}.  They have been obtained  under a bias tension 
of -2~V, and with a \CoF\ source to perform energy calibration (Fig.~\ref{fig:Calibr}).
Though they are satisfactory, in 
the heat-ionisation plane, the 122~keV peak is somewhat smeared.  This is due 
to the incomplete charge collection of a minority of events in all 
likelihood near the surface of the detector, and will be discussed 
later on in this paper.
\begin{table}[b]
	\centering
	\caption{Detector performance under a -2~V bias tension.  
	Ionisation energies are equivalent electron keV.}
	\begin{tabular}{lcc}
		\hline
		 & Heat & Ionisation  \\
		\hline
		FWHM resolution at 122 keV & 1.2 & 1.25  \\
		FWHM baseline noise & 0.85 & 1.1  \\
		Effective energy threshold & 4 & 4  \\
		\hline
	\end{tabular}
	\label{tab:DetectorPerformance}
\end{table}

The distribution of gamma-induced recoil events has been studied further by means of a 
\CoS\ 
source.  This isotope has the advantage over \CoF\ of emitting higher 
energy photons, less susceptible to surface effects in the detector, 
and providing a Compton-scattering plateau.  
With such a source, around 10\% of events suffer from incomplete 
charge collection.
Ignoring for the time being these pathological cases, the 
spread of the photon events, obtained by studying the ionisation to 
heat ratio in the 5 to 200~keV heat range, is compatible with the 
assumptions that, over the studied energy scale, the resolutions of 
both channels are constant, the channel fluctuations are uncorrelated, and that at 
a given energy the dispersion of each channel is a gaussian variable.  A 
slight systematic overestimation of the ionisation at low energies is 
under investigation.

The distribution of neutron-induced recoil events in heat-ionisation space is obtained 
using a \Cf\ source, as done previously \cite{art:Shutt,art:Edelw}.  
Less 
statistics are available than with the photon source; moreover, 
because of the forward scattering of the neutrons on heavy nuclei 
which lowers the recoil energies, this study 
only covers the 5 to 50~keV heat interval.  
The spread of the ionisation 
over heat distribution shows a widening as heat increases.  This is 
being compared to theoretical predictions \cite{art:Lindhard}.  The 
neutron zone that will be of use in Section~\ref{sec:PhysicsAnalysis} is defined in 
ionisation over heat space, and for each heat bin, as being within three 
standard deviations of the mean of the distribution.

\begin{figure}[ht]
	\centering
	\epsfig{file=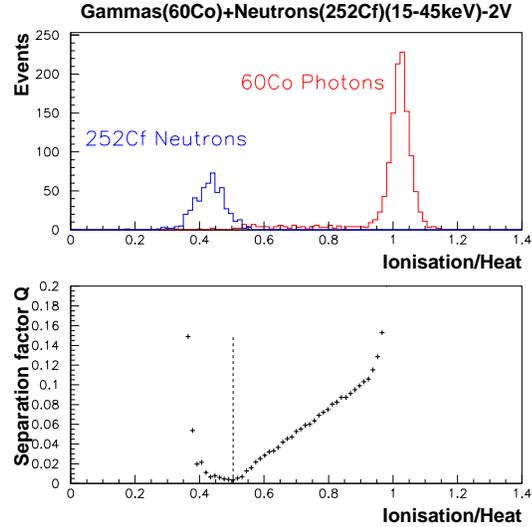,width=\linewidth}
	\caption{Best-case separation of electron recoils from nuclear recoils in the 
15-45~keV heat range, under -2~V of bias tension.  Top graph shows 
distribution of the ionisation to heat ratio for two calibration 
populations of photons and neutrons, from a \CoS\ and a \Cf\ source 
respectively.  Bottom graph presents evolution of separation quality 
factor $Q$ as a function of the aforementioned ratio; vertical dashed 
line marks minimum.}
	\label{fig:BestQ}
\end{figure}

Using the \Cf\ and \CoS\ calibrations, it is possible to obtain a best 
case rejection factor for gammas and acceptance of neutron-type events in the detector.
Over the 15 to 45~keV heat spread, the rejection of gammas 
is $99.7\pm0.2\%$ for an acceptance of neutrons amounting to $94\pm6\%$, 
which yields a 
separation quality factor $Q$ (defined as per Ref.~\cite{art:Gaitskell}) 
of $0.004\pm0.002$, 
as shown in 
Fig.~\ref{fig:BestQ}. 
Though impressive, it must be stressed that this is a best-case result only, 
since, in practice, current real data runs contain proportionately more low 
energy events which are more prone to suffer from surface effects in the detector.  
The ensuing incomplete charge collection pollutes the population 
divide as will be observed below.

\section{PHYSICS RESULTS}
\label{sec:PhysicsAnalysis}
\begin{figure}[htb]
	\centering
	\epsfig{file=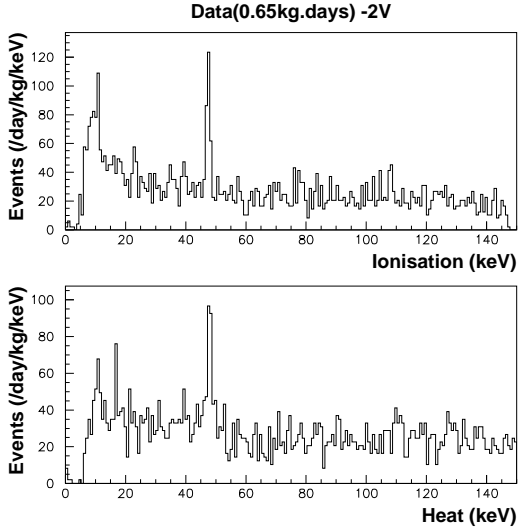,width=\linewidth}
	\caption{0.65~kg.days of raw data taken at 
	a 2~V reverse  bias 
tension.  The 47~keV peak of \Pb\ stands out.  Rate over 15-45~keV range 
is approximately 30~events/day/kg/keV.}
	\label{fig:HeatSpectrum}
\end{figure}

\begin{figure}[htb]																										
	\centering
	\epsfig{file=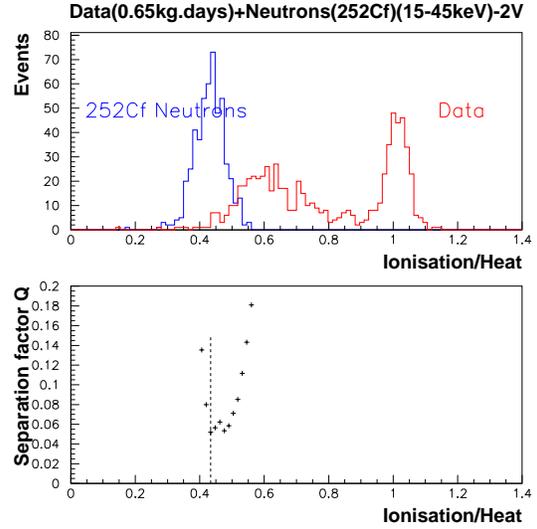,width=\linewidth}
	\caption{Real-life distribution	of electron	and	nuclear	recoils	in
	ionisation over	heat space,	for	the	15-45~keV energy range.	Top shows that background data contains a population	of
	events falling between calibration neutrons	and	well behaved electron
	recoils.  Bottom	demonstrates further that populations have melded compared
	to the best-case scenario presented	in Fig.~\ref{fig:BestQ}.}
	\label{fig:TrueQ}
\end{figure}

Nearly twenty different physics runs taken over late July and early 
August have been compiled, amounting to 0.65~kg.days.  For these runs, 
a 2~V reverse bias tension was applied to the detector, and only minimal 
radioactive shielding was in place: neither Roman lead close shield, 
nor any form of radon removal.  An unsurprisingly high raw rate of 
around 30~events/day/kg/keV, is 
therefore obtained in the 15 to 45~keV range as shown in 
Fig.~\ref{fig:HeatSpectrum}.  Moreover, summary 
inspection of the data shows a gamma peak at approximately 47~keV.  
This can be attributed to photons from \Pb\ pollution.  
Furthermore, a cursory glance at the distribution 
(top of Fig.~\ref{fig:TrueQ}) reveals that separation of nuclear and electron recoils has 
regressed with respect to the calibrations.
	 Indeed, the distribution														
of the ionisation over heat	ratio now clearly shows	a third																
population,	straddling the photons and the expected position	of the													
neutrons.	
	 This reflects the																
incomplete charge collection of	many low energy	electron recoils which											
therefore migrate towards the neutron zone.

To circumvent this,
 a
 cut is made, keeping only the lower half of the neutron 
zone in ionisation over heat space as	defined	in the previous	section.  
The resulting spectrum, adjusted 
for the neutron acceptance of 0.5 
now,
 leads to a rate of 
about 1~event/day/kg/keV and an upper limit at a 90\% confidence level 
of 1.6~events/day/kg/keV.  This encouraging result remains undoubtedly overly 
cautious, as it assumes that none of the events in the neutron zone 
are due to poor charge collection of electron recoils.

Indeed if the shape of the background was known, it could be subtracted 
from the data and an optimal cut could be 
determined~\cite{art:Gaitskell}.
However, as Fig.~\ref{fig:TrueQ} confirms, the \CoS\ data clearly does not reflect 
the true background.  Thus the 
data is itself used under the unverifiable assumption that it contains 
no signal. 
This affords less discrimination than possible in the best case, with 
an acceptance of $50\pm4\%$, a rejection factor of $98.7\pm0.4\%$ 
and a separation factor $Q = 0.05\pm0.02$.  
These entail, 
{\em still under the assumption that the experimental data contains no 
signal}, 
an upper limit following~\cite{art:Gaitskell} at the 90\% conf. level of 
0.5~events/day/kg/keV for the 15-45~keV range. 
This includes a contribution from the statistical uncertainty of the neutron 
calibration amounting to  0.04~events/day/kg/keV.  
The discrepancy 
between 
the result cited in the previous paragraph and that just mentioned is 
linked to the assumption made in the latter.  In the future, strongly increased 
statistics should enable a more adequate method of analysis based on a 
weaker hypothesis.

\section{SURFACE EFFECTS AND INFLUENCE OF BIAS TENSION}
Despite the promising results just demonstrated, off-axis events 
evidently
hinder 
the rejection capabilities of this 
type of detector.
Incomplete charge collection would account for events displaying only 
ionisation deficiency.  Incomplete charge collection coupled to the 
Luke-Neganov effect (enhancement of heat signal by the process of 
collecting
charges~\cite{art:Luke}) can explain the non-horizontal drooling observed 
for instance around the 122~keV photoelectric peak of \CoF\ (Fig.~\ref{fig:Calibr}). These 
effects are expected to affect mainly events occurring near the 
surface of the detector, and can be 
described as loss of part of the back-diffused charge when it nears 
the surface~\cite{art:Penn,art:Shutt2}.  The third population, lying between 
neutrons and gammas in the ionisation over heat distribution, could 
thus be understood as being superficial events, in all likelihood 
betas, or low energy X-rays or gammas.  Identification and elimination of the 
sources of these low energy events is a priority. That said 
events manage to make it to the detector implies they may originate from 
its 
immediate vicinity. 
As a matter of fact, the 
NTD thermometer is itself a prime 
suspect.
In addition, work is underway to further study 
effects on the off-axis events of bias tensions down to $-12$~V, 
as there is some hope of improvement 
since a stronger electric field might be more efficient in collecting charges 
deposited near the surface~\cite{art:Shutt2}.  
Such high bias tensions are acceptable in the current detector 
because of its p-i-n design.

\section{PERSPECTIVES FOR EDELWEISS}
Analysis work will continue on the -2~V bias tension data as well as on 
1.25~kg.days of more recent -6~V 
data.  Regarding the 
current experimental setup, much effort is now being put into reducing 
its radioactive background.  This includes investigating all surfaces 
close to the detector as well as the bolometer itself, and in 
particular the NTD thermometer.  
Eliminating to a great extent any background from the 
latter will be possible shortly by switching to resistive thermometers 
made of $\mbox{NbSi}$ thin-films~\cite{art:Marnieros}.
Moreover, two additional bolometers of 
a design similar to that of the 
current one should be installed in the near term at the LSM, followed 
by three 350 g scaled up devices.
Lastly, planning is underway for the cryogenic infrastructure and 
acquisition logistics of a 10 kg array of detectors.

\section*{ACKNOWLEDGEMENTS}
The collaboration as a whole wish to thank the staff of the LSM for their 
efficient
support.  The speaker is also grateful to J. Mallet and 
J. 
Rich (DAPNIA/SPP) for 
timely and useful discussion.

\end{document}